\documentclass{fic-l}
\usepackage{graphicx}
\usepackage{psfrag}
\usepackage{amsmath}
\usepackage{amssymb}
\usepackage{amsfonts}

\renewcommand{\subjclassname}{%
  \textup{2000} Mathematics Subject Classification}

\theoremstyle{definition}

\theoremstyle{remark}

\numberwithin{equation}{section}
\newcommand{\ele}[3]{\langle #1 \mid #2\mid #3\rangle}
\newcommand{\be}{\begin{equation}}
\newcommand{\ee}{\end{equation}}
\newcommand{\ket}[1]{|#1\rangle}

\newcommand{\ba}{\begin{eqnarray}}
\newcommand{\ea}{\end{eqnarray}}
\newcommand{\bary}{\begin{array}}
\newcommand{\ear}{\end{array}}

\begin{document}
\title[Localization and superselection rules in pyramidal molecules]{Environment induced localization and superselection rules \\ 
\ in a gas of pyramidal molecules}
\author{Giovanni Jona-Lasinio}
\address{Dipartimento di Fisica and INFN \\
Universit\`a di Roma ``La Sapienza''\\
Piazzale A. Moro 2, Roma 00185, Italy}
\email{gianni.jona@roma1.infn.it}
\author{Carlo Presilla}
\address{Dipartimento di Fisica and INFM\\
Universit\`a di Roma ``La Sapienza''\\
Piazzale A. Moro 2, Roma 00185, Italy}
\email{carlo.presilla@roma1.infn.it}
\author{Cristina Toninelli}
\address{Dipartimento di Fisica, Universit\`a di Roma ``La Sapienza'',\\
Piazzale A. Moro 2, Roma 00185, Italy}
\email{cristina.toninelli@roma1.infn.it}
\thanks{Work supported in part by Cofinanziamento MURST protocollo MM02263577\_001}

\date{\today}

\begin{abstract}
We propose a model to describe a gas of pyramidal molecules interacting 
via dipole-dipole interactions. A cooperative effect induced by the 
interaction  modifies the tunneling properties between the classical 
equilibrium configurations of the single  molecule. 
The model suggests that, for sufficiently high gas density, the molecules
become localized in these classical configurations. 
On this basis it is possible to  explain the shift and the disappearance 
of the inversion line observed upon increase of the pressure in a gas of 
ammonia or deuterated ammonia. 
The same mechanism also accounts for the presence of stable optical 
activity of certain pyramidal molecules.
We discuss the concept of environment induced superselection rule which 
has been invoked in connection with this problem.
\end{abstract}
\maketitle

\section{Introduction}
\label{Introduction}

In this work we consider the behavior of gases of pyramidal molecules, 
i.e. molecules of the kind $XY_3$ like for example ammonia $NH_3$, 
deuterated ammonia $ND_3$, phosphine $PH_3$ and arsine $AsH_3$. 
We begin by presenting an outline of the physical problem and of its 
history that dates back to the early developments of quantum mechanics.

For the single pyramidal molecule, owing to the great differences 
of characteristic energies and times, 
we can perform some adiabatic approximations. 
The electronic motion can be separated from the nuclear one 
(Born-Oppenheimer approximation). 
We can further separate the one-dimensional motion of inversion 
of the nucleus $X$ across the plane containing the three nuclei $Y$. 
The form of the effective potential for this motion is 
a double well which is symmetric with respect to the inversion plane 
\cite{Townes,Weston}.
The corresponding non degenerate eigenstates must be symmetric or 
anti-symmetric 
with respect to the inversion plane. 
Due to tunneling across the finite potential barrier, these eigenstates 
are delocalized in the two minima and, for energies below the barrier 
height, 
are grouped in doublets, i.e. couples of states with a relative splitting 
in energy small in comparison with the distance from the rest of the 
spectrum. 
For the pyramidal molecules under consideration,
the thermal energy $k_BT$ at room temperature is much smaller 
than the distance between the first and the second doublet so that  
we can reduce the 
problem to the study of a two-level system corresponding to the
symmetric and anti-symmetric states of the first doublet. 
 
The existence of delocalized stationary states is clearly in 
disagreement with the usual chemical view which, 
relying upon classical theory, considers the molecules as objects 
with a well defined spatial structure. 
In particular, for the molecules under consideration 
the classical view predicts one of the two pyramidal configurations 
corresponding to the nucleus $X$ localized in one of the wells of the 
inversion potential. 

The quantum prediction of stationary delocalized states implies 
the presence of a line in the absorption spectrum, the so called 
inversion line, at a frequency $\bar\nu = h^{-1}\Delta E$, 
where $\Delta E$ is
the energy splitting of the first doublet. 
The experimental results are strongly dependent on the 
different kind of atoms that form the pyramidal molecule. 
Experiments performed with $NH_3$ and $ND_3$ reveal the existence of 
an inversion line, while this line is not found in the case of $AsH_3$ 
and $PH_3$.
These results cannot be considered a demonstration of the 
existence of degenerate classical states for arsine and phosphine.
The values of the energy splitting for these molecules is so small, 
see Table \ref{dati}, that the inversion line, even if it existed, 
would not be detectable. 
However, there is a stronger evidence suggesting the existence of 
stationary localized states, at least for arsine
(similar experiments are, as far as we know, not available in the case 
of phosphine).
It has been found that some molecules of the kind $AsYWZ$,
i.e. molecules obtained by substituting hydrogens with different 
atoms, show stable optical activity \cite{Constain}. 
This means that the stationary states of the $AsHWZ$ molecule are chiral,
i.e. they cannot be superimposed on their mirror image by a rotation, 
in agreement 
with the classical prediction of localized pyramidal states.
The discrepancy between the presence of optical activity for some kind
of molecules and the quantum prediction of delocalized stationary states 
was pointed out for the first time in \cite{Hund} 
and is usually referred to as Hund paradox or chirality problem.
\begin{table}
\begin{center}
\begin{tabular}{l|c|c|c|c}
 & $NH_3$ & $ND_3$ & $PH_3$ & $AsH_3$\\
\hline
$\Delta E$ (cm$^{-1}$)&$0.81$&$0.53\times 10^{-1}$&$3.34\times 10^{-14}$&$2.65\times 10^{-18}$\\
\hline
$\mu$ (Debye)&1.47&1.47&0.58&0.20\\
\hline
\end{tabular}
\end{center}\caption{
Energy splitting of the first doublet $\Delta E$ and
electric dipole moment $\mu$ for the molecules
$NH_3$, $ND_3$, $PH_3$, $AsH_3$ \cite{Weston}.
We remind that 1 cm$^{-1}$ = $1.984 \times 10^{-23}$ J and 
1 Debye = $3.33 \times 10^{-30}$ Cm.}
\label{dati}
\end{table}

The frequency $\bar\nu$ of the inversion line measured in $NH_3$ and 
$ND_3$ gases shows a dependence on the gas pressure $P$
\cite{Penrose1,Penrose2,Birnbaum}.
Starting from the expected value $h^{-1} \Delta E$ at $P \simeq 0$,  
$\bar\nu(P)$ decreases by increasing $P$ and vanishes at a critical 
pressure 
$P_{cr} \simeq 2$ atm for $NH_3$ and $P_{cr} \simeq 0.1$ atm for $ND_3$. 
No quantitative theory has been proposed so far for this phenomenon.

As early as 1949, in a short qualitative paper \cite{Anderson}
Anderson made the hypothesis that dipole-dipole 
interaction may induce a localization of the molecular states.
In this way the important idea was introduced that inter-molecular 
interactions may be responsible for the observed phenomena.
Thirty years later Pfeifer studied a two-level system   
modeling a molecule which has  an almost degenerate ground state and is
coupled to the quantized electromagnetic field \cite{Pfeifer}.
He argued that the coupling of the molecule to the radiation field 
yields two symmetry-breaking effective ground states
belonging to two different sectors of the Hilbert space separated by a 
superselection rule.
These states are localized and one is the mirror image of the other.
Later it was argued that this localization cannot take 
place at finite temperature \cite{Amann,Wightman0,WightmanI}. 

A quantitative discussion of the effects induced by coupling a 
single molecule to the environment constituted by the other 
molecules of the gas was made in \cite{jonaclaverie1,jonaclaverie2}.
In this work it was shown that, due to the instability
of tunneling under weak perturbations \cite{jonamarti1,jonamarti2},
the order of magnitude of the molecular dipole-dipole interaction 
may account for localized ground states in the case of arsine 
and phosphine while delocalized states were predicted for ammonia.
Therefore, the appearance of chirality was interpreted as  
a collective effect in a system of infinitely many degrees of freedom.
This suggests that a phase transition may be invoked to explain the
behavior of different pyramidal molecules or the behavior of the 
same kind of molecule under variation of thermodynamic parameters
like pressure.  

We have implemented this idea by constructing a 
simplified model of a gas of pyramidal molecules which exhibits the
desired properties and allows a direct comparison with experimental
data. 
Our model predicts, for sufficiently high inter-molecular interactions, the presence of two degenerate   ground states corresponding to the  different localizations of the molecules. This means that there is a quantum phase transition for a critical value of density (or pressure in the case of constant temperature).
A study at non zero temperature shows that this transition affects the behaviour of the gas at room temperature and gives a reasonable explanation of the experimental results \cite{Penrose1, Penrose2, Birnbaum}. In particular, it provides a correct description of the 
disappearance of the inversion line of $NH_3$ and $ND_3$ 
on increasing the pressure.

In the present paper, after a description of the model, we discuss its mean field  approximation and the appearance of a degeneracy for the ground state. We then analyze the linear response to an external field at zero temperature to obtain the low energy spectrum and  its dependence on the inter-molecular interaction. This shows that the frequency of the inversion line decreases upon increasing density and vanishes at the critical density for which the ground state becomes degenerate.  
For a treatment at finite temperature and a detailed comparison with experiments  we refer to our paper \cite{JPT}.

In the last section we  discuss  the interpretation of the chirality problem in terms of a superselection rule.

\section {Description of the model}
\label{Model}
As we mentioned in the introduction, for an isolated molecule $XY_3$
we can adiabatically separate the electronic motions from the nuclear 
ones.
The latter essentially reduce to the rotational motion of the molecule, 
the vibrations of the nuclei around their equilibrium
positions and the inversion motion of the nucleus $X$ across the plane 
of the nuclei $Y$. 
At room temperature the rotational and vibrational degrees of freedom are 
much faster then the inversion motion and can be taken into account 
through a renormalization of the potential barrier for the inversion 
motion.
We can further reduce the 
inversion dynamics to that of a two-level system corresponding to the 
ground and first excited eigenstates of the corresponding
symmetric double well potential. 
For a general reference see \cite{Townes}.

Consider now a gas of interacting molecules.
If the motion of the centers of mass of the molecules is slow 
in comparison with the two-level dynamics of each isolated molecule,
the translational degrees of freedom can be considered frozen.
In this case we can reduce our model to a lattice 
of unit length $\ell = \varrho^{-1/3}$, 
where $\varrho$ is the gas density,
with a two-state system at each site.
By evaluating the inversion time $\hbar \Delta E^{-1}$ with the data  
of Table \ref{dati}, we see that the lattice approximation 
is reasonable in the case of $NH_3$ and $ND_3$ up to room temperature.  
This approximation has been discussed in general \cite{Lebowitz} 
and we refer to this paper also for references to previous related works.

In order to define a suitable Hamiltonian for our lattice model,
let us begin with the isolated two-level system.
We choose the Hamiltonian operator on the two-state Hilbert space to be
\begin{equation}
-\frac{\Delta E}{2} \sigma^x,
\end{equation} 
where $\sigma^x$ is the Pauli matrix
\begin{equation}
\sigma^x=
\left(
\bary{cc}
0&1\\
1&0\\
\ear
\right).
\end{equation}
The corresponding eigenvectors are 
\begin{eqnarray}
\ket{1} &=& \frac{1}{\sqrt 2}\left(
\begin{array}{c}
 1\\ 1
\end{array}
\right)
\label{1}
\\
\ket{2} &=& \frac{1}{\sqrt 2} \left(
\begin{array}{c}
1\\-1
\end{array}
\right),
\label{2}
\end{eqnarray}
with eigenvalues $-\Delta E/2$ and $+\Delta E/2$, respectively.
For $N$ independent two-state systems, the Hamiltonian is 
\begin{equation}
H_0=\sum_{i=1}^N-\frac{ \Delta E}{2}
1_1\otimes 1_2\otimes\ldots\otimes\sigma^x_i\otimes\ldots 1_N,
\label{libera}
\end{equation}
where the subscript indicates the single particle space to which 
the operators are applied.

In order to modeling the interaction among the molecules we briefly 
recall the discussion made in \cite{jonaclaverie1}.
If we consider the whole gas it is possible that, owing to some 
fluctuation in the state of the system, a single molecule becomes 
localized. This molecule consequently acquires an electric 
dipole moment $\mbox{\boldmath$\mu$}$ that polarizes the surrounding gas. 
The latter, in its turn, produces an electric reaction field 
$\mbox{\boldmath$E$}_R$, collinear with $\mbox{\boldmath$\mu$}$,
so that the interaction 
$-\mbox{\boldmath$\mu$} \cdot \mbox{\boldmath$E$}_R$
tends to stabilize the localized state of the initial molecule.
In other words, there appears a cooperative effect which tends to 
stabilize the localization.
To mimic this effect in the Hilbert space where $H_0$ acts, 
we choose the interaction Hamiltonian
\begin{equation}
H_{int}=\sum_{i=1}^{N} \sum_{j=i+1} ^{N}
 g(i,j)\, 1_1\otimes\ldots\otimes\sigma^z_i\otimes\ldots
\otimes\sigma^z_j\otimes\ldots
\otimes1_N,
\label{interazione}
\end{equation}
where $g(i,j)=g(j,i)<0$ and $\sigma^z$ is the Pauli matrix 
\begin{equation}
\sigma^z= \left(
\bary{cc}
1&0\\
0&-1\\
\ear \right)
\end{equation}
whose eigenvectors are the localized states 
\begin{eqnarray}
\label{RRR}
\ket{L}&=&\left(
\begin{array}{c}
1\\0
\end {array}
\right) = \frac{1}{\sqrt{2}} \left( \ket{1} + \ket{2} \right)
\\
\ket{R}&=&\left(
\begin{array}{c}
0\\1
\end {array}
\right) = \frac{1}{\sqrt{2}} \left( \ket{1} - \ket{2} \right),
\end{eqnarray}
with eigenvalues $+1$ and $-1$, respectively.
The total Hamiltonian 
\begin{equation}
H=H_0+H_{int}
\label{modelli}
\end{equation}
contains two competing terms with $H_{int}$ which tends to favor
stationary localized states while $H_0$  favors delocalization.
According to the classical dipole-dipole interaction formula,
we expect $g(i,j)$ to be proportional to the square of the electric 
dipole moment $\mu$ of the molecules and to decrease with the cube
of the distance
\begin{equation}
|g(i,j)| \propto \frac{\mu^2}{\ell^3 |i-j|^3} .
\label{gmul}
\end{equation}

In the context of a mean field approximation, it is useful 
to introduce the parameter
\begin{equation}
G \equiv - 
\sum_{j=1}^{N_c} g(i,j),
\label{JJ}
\end{equation}
wher $N_c$ is a cut-off.
By neglecting border effects and using (\ref{gmul}), we write
\begin{equation}
G = C \frac{\mu^2}{\ell^3},
\label{GP}
\end{equation}
where $C$ is a positive constant which
will  be considered as a phenomenological parameter
to be deduced, in the case of $NH_3$ and $ND_3$, from comparison 
with the experimental data on the inversion line.

\section{Ground state in the mean field approximation}
\label{Transition}

The mean field approximation for the ground state
corresponds to minimizing the energy functional 
$\ele{\psi}{H}{\psi}$ within the class of states of the form
\begin{equation}
\ket{\psi} =\ket{\lambda}\ldots\ket{\lambda}
\label{psi},
\end{equation}
where
\begin{equation}
\ket{\lambda}=a\ket{1}+b\ket{2}
\label{coeffi}
\end{equation}
and
\begin{equation}
{\mid a\mid}^2+{\mid b\mid}^2=1.
\label{cane}
\end{equation}
This problem reduces to the solution of the non linear eigenvalue equation
\begin{equation}
h(\lambda)\ket{\lambda}=\eta(\lambda) \ket{\lambda},
\label{auto}
\end{equation}
where $h(\lambda)$ is the self-consistent single particle 
Hamiltonian
\begin{equation}
h(\lambda)=-\frac{\Delta E}{2}\sigma^x-
G\sigma^z\ele{\lambda}{\sigma^z}{\lambda}.
\label{acca}
\end{equation}
The non linear term in (\ref{acca})
accounts for the interaction of a single
molecule with the reaction field and represents the polarization 
energy of the gas.

We find $\ket{\lambda}$ by solving Eq. (\ref{auto}) and evaluate the 
energy functional
$\ele{\psi}{H}{\psi}$ with the corresponding product states (\ref{psi}).
For the states which minimize $\ele{\psi}{H}{\psi}$
we obtain the following results.
If $G< \Delta E/2$ there is only one mean-field ground state
\begin{equation}
\ket{\psi_0}=\ket{1}\ldots\ket{1},
\end{equation}
with energy
\begin{equation}
E_0=\ele{\psi_0}{H}{\psi_0}=-N \frac{\Delta E}{2}.
\end{equation}
If $ G \geq \Delta E/ 2 $ there are two degenerate 
mean-field ground states
\ba
\ket{\psi_0^L}&=&\ket{\lambda^L}\ldots\ket{\lambda^L}\\
\ket{\psi_0^R}&=&\ket{\lambda^R}\ldots\ket{\lambda^R},
\label{ground}
\ea
where
\ba
\ket{\lambda^L}&=&\sqrt{\frac{1}{2} +\frac{\Delta E}{4G}} 
\ket{1}+\sqrt{\frac{1}{2}-\frac{\Delta E}{4G}}\ket{2} 
\label{chiral1}\\
\ket{\lambda^R}&=&\sigma^x\ket{\lambda^L},
\label{chiral2}
\ea
with energy
\begin{equation}
E_0=-N\frac{\Delta E}{2}-\frac{N}{2G}
\left( \frac{\Delta E}{2}-G \right)^2.
\end{equation}

By defining the critical value
\begin{equation}
G_{cr}=\frac{\Delta E}{2}, 
\end{equation}
we distinguish the following two cases.
For $G \in (0,G_{cr})$ 
the ground state of the system is approximated by a 
condensate  of delocalized symmetric single particle states 
corresponding to the ground state of an isolated molecule.
For $G \in (G_{cr},\infty)$ we have two different condensates which approximate
the ground state of the system. The corresponding single particle states  transform 
one into the other under the action of the inversion operator $\sigma^x$, 
as shown by Eq. (\ref{chiral2}), and, for $G\gg G_{cr}$, they  become localized
\ba
\lim_{\frac{\Delta E}{G}\rightarrow 0}\ket{\lambda^L}&=&\ket{L}
\\
\lim_{\frac{\Delta E}{G}\rightarrow 0}\ket{\lambda^R}&=&\ket{R}.
\ea

The above results suggest the existence of a quantum phase transition  
at $G=G_{cr}$. 
According to Eq. (\ref{GP}), this phase transition can be obtained
by increasing the density $\varrho$ of the gas above the critical value 
\begin{equation}
\varrho_{cr}=\frac{\Delta E}{2}\frac {1}{\mu^2 C}.
\label{pcr}
\end{equation}
In the next Section we will discuss the behavior of the inversion
line obtained in the range $0<\varrho<\varrho_{cr}$ when the gas,
still being in the region characterized by delocalization, approaches the phase transition.

\section{Absorption spectrum}
\label{Spectrum}

When the gas is exposed to an electro-magnetic radiation of
frequency $\nu_0=\omega_0/2\pi$, we add to the Hamiltonian 
(\ref{modelli}) the perturbation
\begin{equation}
H_{em}(t)=\epsilon f(t) S
\end{equation}
where $\epsilon$ is a small parameter,
\begin{equation}
f(t)=\theta(t) \cos(\omega_0 t),
\end{equation}
$\theta(t)$ being the Heaviside function, 
and $S$ is a time independent operator.
In the long-wavelength approximation, $\lambda_0 \gg d$, where $d$
is the molecule size, we have
\begin{equation}
S = \sum_{i=1}^{N} 1_1\otimes 
\ldots\otimes\sigma^z_i\otimes \ldots\otimes 1_N.
\label{em}
\end{equation} 

At zero temperature the absorption spectrum induced by the electro-magnetic perturbation
consists in a series of lines of frequency 
$\nu_n = (E_n-E_0) h^{-1}$ corresponding to transitions of the
system from the ground state of $H$ to the $n$-th excited level.
The corresponding transition probability is proportional to the 
square of the matrix element 
\begin{equation}
M_{0n}\equiv  \ele{\psi_0}{S}{\psi_n}.
\end{equation}
Since we are not able to determine exactly the eigenvalue problem 
$H\psi_n=E_n \psi_n$ for the $N$ particle Hamiltonian (\ref{modelli}), 
we will find an approximate mean-field solution in the framework of 
the linear response theory.
 
Let us define the function
\begin{equation}
{\mathcal S} (t)\equiv\ele{\psi (t)}{S}{\psi (t)},
\label{roro}
\end{equation}
where 
$\ket{\psi(t)}= \exp[-i(H+H_{em})t/\hbar] \ket{\psi_0}$.
By expanding ${\mathcal S}(t)$ in powers of $\epsilon$, 
for the first order term ${\mathcal S}_1(t)$ we find \cite{Blaizot}
\begin{equation}
\label{pooo}
\tilde{{\mathcal S}}_1(\omega)=\tilde{f}(\omega){\mathcal R}(\omega),
\end{equation}
where $\tilde{{\mathcal S}}_1(\omega)$ and $\tilde{f}(\omega)$ are the 
Fourier transforms of ${\mathcal S}_1(t)$ and $f(t)$ and 
\ba
{\mathcal R}(\omega) = \sum_n \left| M_{0n} \right|^2 
\left[ \frac{1}{\hbar\omega-(E_n-E_0)+i\delta} 
- \frac{1}{\hbar\omega+(E_n-E_0)+i\delta} \right] 
\label{poip}
\ea
with $\delta \to 0^+$.
The frequencies of the spectroscopic lines and the associated
transition probabilities are therefore given by
the poles and the residues of ${\mathcal R}(\omega)$.
In order to determine ${\mathcal R}(\omega)$, we evaluate $\tilde{{\mathcal S}}_1(\omega)$
in a mean-field approximation and then use (\ref{pooo}).

The mean-field approximation for the time-dependent state 
$\ket{\psi(t)}$ is obtained by extremizing the action functional
\begin{equation}
{\int_0}^T \ele {\psi (t)}{-i\hbar\partial _t+H+H_{em}} {\psi (t)} dt
\end{equation}
in the class of states of the form 
\begin{equation}
\ket{\psi(t)}=\ket{\lambda (t)}\ldots\ket{\lambda (t)}.
\end{equation}
The single-particle state $\ket{\lambda(t)}$ is thus determined by 
the nonlinear Schr\"odinger equation
\begin{equation}
i\hbar\frac{d\ket{\lambda(t)}}{dt}=
\left[ h(\lambda(t)) + \epsilon f(t) \sigma^z \right]
\ket{\lambda(t)},
\label{hmft}
\end{equation}
where $h(\lambda)$ is given by (\ref{acca}).
Let us assume that the gas is in the quantum region
$G \in (0,G_{cr})$, so that the mean-field ground 
state is $\ket{\psi_0}= \ket{1}\ldots\ket{1}$.
Since $\ket{\psi(0)}=\ket{\psi_0}$,
Eq. (\ref{hmft}) has to be solved with the initial condition
\begin{equation}
\ket{\lambda (0)}=\ket{1}.
\end{equation}
The calculation shows  that ${\mathcal R}(\omega)$
has a single pole corresponding to a line in the absorption spectrum 
with frequency
\begin{equation}
\bar\nu=\frac{\Delta E}{h}\sqrt{1-\frac{2G}{\Delta E}}.
\label{cica}
\end{equation}
The residue of ${\mathcal R}(\omega)$ at this pole is
\begin{equation}
\frac{N}{\sqrt{1-\frac{2G}{\Delta E}}}.
\label{residue}
\end{equation} 

\begin{figure}   
\begin{center}   
\psfrag{assex}[][][0.9]{$\varrho/\varrho_{cr}$}
\psfrag{assey}[][][0.9]{$\bar\nu(\varrho)/\bar\nu(0)$}
\includegraphics[width=9cm]{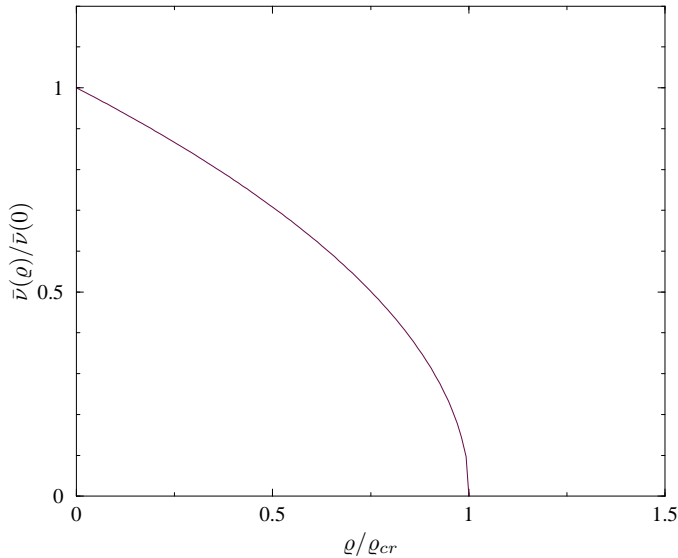}
\caption{Inversion-line frequency $\bar\nu$ 
as a function of the gas density $\varrho$.}
\label{fig1}
\end{center}
\end{figure}

Equation (\ref{cica}) shows a dependence of $\bar\nu$ on the coupling
constant $G$. 
For $G=0$ we have $\bar\nu = \Delta E h^{-1}$, namely
the inversion line frequency expected from the quantum 
theory of an isolated molecule.
In the region $0<G<G_{cr}$,  
$\bar\nu$ decreases by increasing $G$ and vanishes for 
$G=G_{cr}$.  In figure \ref{fig1} we show the described behavior of the inversion line frequency as a function of the gas density $\varrho=G/C\mu^2$.

The residue (\ref{residue}) multiplied by the photon energy 
$h \bar\nu$ provides, up to a constant, the intensity $I$ 
of the inversion line 
\begin{equation}
I  \propto h \bar\nu \frac{N}{\sqrt{1-\frac{2G}{\Delta E}}}
= N \Delta E.
\label{intint}
\end{equation}
Note that the divergence at $G=G_{cr}$ shown by the residue 
is cancelled by the factor $h \bar\nu$ 
which vanishes in the same limit.

\section{Optical activity and superselection rules}
\label{ssr}

Our model does not distinguish molecules of the type $XY_3$ from the 
substituted derivatives $XYWZ$. In fact, we believe that this distinction 
should not matter  for the disappearance of the inversion line.  
However, an important difference between the two cases is that for 
$XY_3$ the localized states can be obtained one from the other either 
by a rotation or by space inversion, while for $XYWZ$ they  can be 
connected only by space inversion. 
This implies that $XYWZ$ molecules, whenever become localized, 
i.e. at a density greater than the critical value, are chiral and 
must show  stable optical activity. 
Stable optical activity is  commonly observed for  some 
substituted derivatives of $AsH_3$ for which the critical density is 
exceedingly small.
For the theory of optical activity we refer, for example, to 
\cite{Caldwell}.

To explain the stability of the optical activity, i.e. why the localized 
states are stable, although they  are not 
eigenstates of the single-molecule Hamiltonian, an environment induced  
superselection rule has been invoked 
\cite{Pfeifer,Amann,Wightman0,WightmanI}.
        
Following \cite{Wightman0,WightmanI} we say that in a system with 
infinitely many degrees of freedom there is a superselection 
rule between two different sectors of a Hilbert space if there are 
neither spontaneous transitions between their state vectors, 
nor any measurable quantity with matrix element different from zero
between them. 
This implies that any coherent superposition of states among the two 
sectors is physically meaningless and that we can construct a classical 
observable 
(i.e. an observable with zero dispersion on all the states) which 
acquires the same value on all the vectors of each sector. 
Examples are the charge and the Boson-Fermion superselection rules. 
The first refers to the impossibility of coherent superpositions of  
states with different charge,
the second  to the impossibility of superposing states with integer 
and half-integer angular momentum.

With respect to this definition, in the limit of an infinite number of 
molecules (thermodynamic limit) our model has a superselection rule  
similar to that found by Pfeifer analyzing a single molecule interacting 
with the radiation field \cite{Pfeifer}. Above the critical density, 
the Hilbert space separates into two sectors generated by the ground 
state vectors given in mean field approximation by (\ref{ground}).  
These sectors, which we call $\mathcal{H}_L$ and $\mathcal{H}_R$,  cannot be connected 
by any operator involving a finite number of degrees of freedom 
(local operator). 
Let us define chirality 
$\chi=\lim_{N \to \infty} \frac{1}{N}\sum_{i=1}^N 1_1\otimes\ldots 
\otimes\sigma^z_i  \otimes \ldots \otimes 1_N $. 
In the limit $N \to \infty$ we have $\chi \psi= \pm\psi$ for $\psi$ 
in $\mathcal{H}_L$ and $\mathcal{H}_R$, respectively.
Optical activity is expected when $\chi \neq 0$.

From the above definition the concept of superselection rule is  
applicable only to pure states. 
A superselection rule, however, puts some restrictions in the 
construction of ensembles (density matrices) because these cannot contain 
mixed terms corresponding to vectors belonging to different sectors of 
the Hilbert space. 
Indeed, at finite temperature, we find that in the mean field 
approximation the free energy of our model is minimized by two 
density matrices which have vanishing matrix elements between the 
two sectors $\mathcal{H}_L$ and $\mathcal{H}_R$ \cite{JPT}. 
Any other admissible density matrix will be a convex combination 
of these two.

A different point of view consists in considering, instead of the 
whole gas, a subsystem composed by a single molecule or a finite number 
of them still interacting with the rest of the gas.   
In this case a superselection rule in the Hilbert space of the 
subsystem can emerge  dynamically. 
By this we mean that there exist projection operators such that 
the off-diagonal contributions of the subsystem density matrix 
vanish when time $t\to \infty$ \cite{Kupsch}. 
This alternative point of view, usually referred to as environment 
induced superselection rule \cite{Zureck}, can be presumably 
implemented in our model if we describe the interaction 
of a molecule with the rest of the gas by means of a Lindblad
equation \cite{Lindblad}.

\vspace{1cm}

It is a pleasure to dedicate this paper to Sergio Doplicher on the 
occasion of his 60th birthday and we thank him for a very interesting discussion on superselection rules.


\begin{thebibliography}{99}


\bibitem{Townes}C.H. Townes and A.L. Schawlow,
\textit{Microwave Spectroscopy} (Mc Graw-Hill, New York, 1955) 
\bibitem{Weston} R.E. Weston, J. Am. Chem. Soc. \textbf{76}, 2645 (1954).
\bibitem{Constain}C. Constain and M. Sutherland,
J. Phys. Chem. \textbf{56}, 321 (1952).
\bibitem{Hund}F. Hund, Z. Phys. \textbf{series 43}, 805 (1927).
\bibitem{Penrose1}B. Bleaney and J.H. Loubster, 
Nature \textbf{161}, 522 (1948).
\bibitem{Penrose2}B. Bleaney and J.H. Loubster, 
Proc. Phys. Soc. (London) \textbf{A63}, 483 (1950).
\bibitem{Birnbaum}G. Birnbaum and A. Maryott, 
Phys. Rev. \textbf{92}, 270 (1953).
\bibitem{Anderson}P.W. Anderson, Phys. Rev. \textbf{75}, 1450 (1949).
\bibitem{Pfeifer}P. Pfeifer in 
\textit{Energy Storage and Redistribution in Molecules}, 
Hinze Editor (Plenum, New York, 1983).
\bibitem{Amann}A. Amann in 
\textit{Large Scale Molecular Systems: Quantum and Stochastic Aspects. 
Beyond the simple molecular picture}, 
W. Gans, A. Blumen, and A. Amann Editors 
(Plenum Press, New York, 1991).
\bibitem{Wightman0}A.S. Wightman, 
\textit{Some comments on the quantum theory of measurement}
in \textit{Probability methods in mathematical physics},
F. Guerra, M. Loffredo, and C. Marchioro Editors 
(World Scientific, Singapore, 1992), p. 411.
\bibitem{WightmanI}A.S. Wightman, 
Il nuovo Cimento \textbf{110 B}, 751 (1995).
\bibitem{jonaclaverie1}P. Claverie and G. Jona-Lasinio, 
Phys. Rev. A \textbf{33}, 2245 (1986). 
\bibitem{jonaclaverie2}G. Jona-Lasinio and P. Claverie, 
Prog. Theor. Phys. Suppl. \textbf{86}, 54 (1986).
\bibitem{jonamarti1}G. Jona-Lasinio, F. Martinelli, and E. Scoppola, 
Phys. Rep. \textbf{77}, 313 (1981).
\bibitem{jonamarti2}G. Jona-Lasinio, F. Martinelli, and E. Scoppola, 
Commun. Math. Phys. \textbf{80}, 223 (1981).
\bibitem{JPT}G. Jona-Lasinio, C. Presilla, C. Toninelli, 
paper in preparation.
\bibitem{Lebowitz}P. de Smedt, P. Nielaba, J.L. Lebowitz, J. Talbot,
and L. Dooms, Phys. Rev. A \textbf{38}, 1381 (1988).
\bibitem{Blaizot}J.P. Blaizot and G. Ripka, 
\textit{Quantum theory of finite systems}
(The MIT Press, Cambridge, Massachusetts, 1986).
\bibitem{Caldwell}D.J. Caldwell and H. Eyring,
\textit{The theory of optical activity} 
(Wiley Interscience, New York, 1971).
\bibitem{Kupsch}J. Kupsch, J. Math. Phys. \textbf{41}, 5945 (2000).
\bibitem{Zureck}W.H. Zureck, Phys. Rev. D \textbf{26}, 1862 (1982).
\bibitem{Lindblad}G. Lindblad, Comm. Math. Phys. \textbf{48}, 119 (1976).

\end{thebibliography}
\end{document}